\documentclass[useAMS,usenatbib,usegraphicx,letterpaper]{mn2e}
\usepackage{aas_macros,epsfig}

\usepackage[totalwidth=480pt,totalheight=680pt,layoutvoffset=0.5cm]{geometry}

\newcommand{\Mearth}{\ensuremath{M_\oplus}}

\title[The Debris Disk of Solar Analogue $\tau$~Ceti]{The Debris Disk of Solar Analogue $\tau$~Ceti: 
{\it Herschel} Observations and Dynamical Simulations of the Proposed Multiplanet System}

\author[S.~M.~Lawler et al.]{
S.~M.~Lawler\thanks{Email:lawler@uvic.ca}$^{1,2}$,
J.~Di~Francesco$^{2,1}$,
G.~M.~Kennedy$^3$,
B.~Sibthorpe$^4$,
M.~Booth$^5$,
\newauthor
B. Vandenbussche$^6$,
B.~C.~Matthews$^{2,1}$,
W.~S.~Holland$^{7,8}$,
J.~Greaves$^9$,
D.~J.~Wilner$^{10}$,
\newauthor
M.~Tuomi$^{11,12}$,
J.~A.~D.~L.~Blommaert$^{13,14,6}$,
B.~L.~de~Vries$^{15,16}$,
C.~Dominik$^{17,18}$,
\newauthor
M.~Fridlund$^{19,20}$,
W.~Gear$^{21}$,
A.~M.~Heras$^{22}$,
R.~Ivison$^{23,24}$,
G.~Olofsson$^{15}$\\
$^1$Department of Physics \& Astronomy, University of Victoria, PO Box 1700, STN CSC, Victoria, BC V8W 2Y2, Canada\\
$^2$National Research Council of Canada, Herzberg Astronomy \& Astrophysics Program, 5071 West Saanich Road., Victoria, BC V9E 2E7, Canada\\
$^3$Institute of Astronomy, Cambridge University, Madingley Road, Cambridge CB3 0HA, UK\\
$^4$SRON Netherlands Institute for Space Research, 9747 AD, Groningen, The Netherlands\\
$^5$Instituto de Astrof\'{i}sica, Pontificia Universidad Cat\'{o}lica de Chile, Vicu\~{n}a Mackenna 4860, 7820436 Macul, Santiago, Chile\\
$^6$Institute of Astronomy KU Leuven, Celestijnenlaan 200D, 3001 Leuven, Belgium\\
$^7$UK Astronomy Technology Centre, Royal Observatory Edinburgh, Blackford Hill, Edinburgh EH9 3HJ, UK\\
$^8$Institute for Astronomy, University of Edinburgh, Royal Observatory Edinburgh, Blackford Hill, Edinburgh EH9 3HJ, UK\\
$^9$SUPA, School of Physics and Astronomy, University of St. Andrews, North Haugh, St. Andrews KY16 9SS, UK\\
$^{10}$Harvard-Smithsonian Center for Astrophysics, 60 Garden Street, Cambridge, MA 02138, USA\\
$^{11}$Centre for Astrophysics Research, University of Hertfordshire, College Lane, AL10 9AB, Hatfield, UK\\
$^{12}$Departamento de Astronomia, Universidad de Chile, Camino del Observatorio 1515, Las Condes, Santiago, Chile\\
$^{13}$Astronomy and Astrophysics Research Group, Department of Physics and Astrophysics, Vrije Universiteit Brussel, Pleinlaan 2, 1050 Brussels, Belgium\\
$^{14}$Flemish Institute for Technological Research (VITO), Boeretang 200, 2400 Mol, Belgium\\
$^{15}$Department of Astronomy, Stockholm University, AlbaNova University Center, 10691 Stockholm, Sweden \\
$^{16}$Stockholm University Astrobiology Centre, SE-106 91 Stockholm, Sweden\\
$^{17}$Anton Pannekoek Institute, University of Amsterdam, Science Park 904, 1098 XH Amsterdam, The Netherlands\\
$^{18}$Department of Astrophysics/IMAPP, Radboud University Nijmegen, PO Box 9010, NL-6500 GL Nijmegen, The Netherlands\\
$^{19}$Institute of Planetary Research, German Aerospace Center, Rutherfordstrasse 2, 124 89, Berlin, Germany\\
$^{20}$Leiden Observatory, University of Leiden, PO Box 9513, 2300 RA, Leiden, The Netherlands\\
$^{21}$School of Physics and Astronomy, Cardiff University, Queens Buildings, The Parade, Cardiff CF24 3AA, UK\\
$^{22}$Scientific Support Office, Science and Robotic Exploration Directorate, ESA/ESTEC, Keplerlaan 1, 2200 AG Noordwijk, The Netherlands\\
$^{23}$European Southern Observatory, Karl Schwarzschild Strasse 2, D-85748 Garching, Germany\\
$^{24}$Institute for Astronomy, University of Edinburgh, Blackford Hill, Edinburgh EH9 3HJ, UK\\
}
\date{Released 2014 Xxxxx XX}

\pagerange{\pageref{firstpage}--\pageref{lastpage}} \pubyear{2014}

\def\LaTeX{L\kern-.36em\raise.3ex\hbox{a}\kern-.15em
    T\kern-.1667em\lower.7ex\hbox{E}\kern-.125emX}

\begin{document}

\label{firstpage}

\maketitle

\begin{abstract}

$\tau$ Ceti is a nearby, mature G-type star very similar to our Sun, with a massive 
Kuiper Belt analogue \citep{Greavesetal2004} and possible multiplanet system 
\citep{Tuomietal2013} that has been compared to our Solar System. 
We present {\it Herschel Space Observatory} images of the debris disk,
finding the disk is resolved at 70~$\mu$m and 160~$\mu$m,
and marginally resolved at 250~$\mu$m.
The {\it Herschel} images and infrared photometry from the literature are 
best modelled using a wide dust annulus with an inner edge between 1-10~AU
and an outer edge at $\sim$55~AU, 
inclined from face-on by $35^{\circ}\pm10^{\circ}$, and with no significant azimuthal structure. 
We model the proposed tightly-packed planetary system of five super-Earths and find that 
the innermost dynamically stable disk orbits are consistent with the inner edge found 
by the observations. 
The photometric modelling, however, cannot rule out a disk inner edge as close to the star as 1 AU, 
though larger distances produce a better fit to the data. 
Dynamical modelling shows that the 5 planet system is stable with the addition of a 
Neptune or smaller mass planet on an orbit outside 5 AU, where the Tuomi et al.\ analysis 
would not have detected a planet of this mass.

\end{abstract}

\begin{keywords}
circumstellar matter $-$ planet-disc interactions $-$ planets and satellites: dynamical evolution and stability $-$ stars: individual: $\tau$~Ceti
\end{keywords}

\section{Introduction}

Although hundreds of planetary systems are now known, we are still trying to 
understand whether or not our Solar System is typical.
The distributions of known planetary system parameters are strongly 
affected by observational biases that are not easy to disentangle from the 
true distributions.
Moreover, our Solar System's architecture (small rocky inner planets, large gaseous outer planets, 
and an outer debris disk) has not yet been found in other systems, most likely due to these same biases.
For example, long time baselines are required to discover planets at greater 
than a few AU
by either the transit or radial velocity (RV) techniques,
and directly imaging planets around mature stars 
like $\tau$~Ceti \citep[5.8~Gyr;][]{MamajekHillenbrand2008},
is difficult
due to the low fluxes of planets after they lose most of their initial heat
from formation \citep[e.g.][]{SpiegelBurrows2012}.

Fortunately, structures in debris disks can indicate the presence of additional planets.
Indeed, one planet so far has been predicted based on disk morphology and then
later discovered by direct imaging or another technique, \citep[$\beta$~Pic~b][]{Mouilletetal1997,Lagrangeetal2010}.
In this paper, we use the debris disk to probe the planetary system around $\tau$~Ceti.

$\tau$~Ceti is a solar-type analogue located only 3.65~pc from the Sun.
The infrared excess toward $\tau$~Ceti has been known for nearly three
decades, first discovered by IRAS \citep{Aumann1985} and later
confirmed by ISO \citep{Habingetal2001}.

\citet{Greavesetal2004}, using the Submillimeter Common-User Bolometer Array
\citep[SCUBA;][]{Hollandetal1999} instrument on the James Clerk Maxwell Telescope (JCMT),
found $\tau$~Ceti to have a significant excess and moderately resolved disk
at 850~$\mu$m, extending 55~AU from the star,
and inferred to be misaligned with the rotational axis of the star.
They fit the observed excess between 60~$\mu$m and 850~$\mu$m with a 
single temperature blackbody at 60~K,
and obtained a disk mass of 1.2~$\Mearth$, about an order of magnitude higher than 
our Kuiper Belt.

Here, we revisit the $\tau$~Ceti disk with higher-resolution far-IR
images taken by the {\it Herschel Space Observatory}\footnote{{\it Herschel} is an ESA space
  observatory with science instruments provided by European-led Principal Investigator
  consortia and with important participation from NASA.},
attempting to better constrain the properties of the disk.
Additionally, we find the observed disk probably does not overlap with the orbits of the 
proposed multiplanet
system \citep{Tuomietal2013}, though we cannot rule out a disk inner edge inside the orbit of the
outermost planet. 

In Section~\ref{sec:obs}, we present the {\it Herschel} observations.
Section~\ref{sec:disk} discusses the constraints these observations
place on the properties of the $\tau$~Ceti debris disk.
In Section~\ref{sec:planets}, we show that the disk inner edge inferred
from the modelling is compatible with the proposed compact multiplanet system,
and we use dynamical simulations to investigate system stability and 
the possible presence of additional planets.
In Section~\ref{sec:disc} we discuss the $\tau$~Ceti disk-planet system
in the context of other known solar systems,
and a summary of our conclusions is given in Section~\ref{sec:concl}.

\section{{\it Herschel} Observations} \label{sec:obs}

\begin{table}
\caption{{\it Herschel} Observations of $\tau$~Ceti} 
\label{tab:obs}
\begin{tabular}{l|lll}
\hline
ObsID & Date & Instrument & Duration (s) \\ \hline
1342199389 & 2010 June 29 & SPIRE 250/350/500 & 2906 \\
1342213575 & 2011 January 31 & PACS 70/160 & 5478 \\
1342213576 & 2011 January 31 & PACS 70/160 & 5478 \\
\end{tabular}
\end{table}

$\tau$~Ceti and its surroundings were observed with the {\it Herschel Space Observatory}
\citep{Pilbrattetal2010} using both the Photodetector Array Camera and Spectrometer 
\citep[PACS;][]{Poglitschetal2010} and the Spectral and Photometric Imaging Receiver
\citep[SPIRE;][]{Griffinetal2010} as part of the Guaranteed Time Key Programme
``Stellar Disk Evolution'' to study the six most well-known debris disks 
(PI: G.~Olofsson; Proposal ID: KPGT\_golofs01\_1). 
Data at 70 $\mu$m and 160 $\mu$m 
were obtained simultaneously using the PACS large scan-map mode 
on 31 January 2011 over a successive scan and a cross-scan each 
lasting 91.3 minutes (ObsIDs: 1342213575 and 1342213576).  The PACS scan speed 
was 20$^{\prime\prime}$ s$^{-1}$.  Data at 250~$\mu$m, 350~$\mu$m, and 
500~$\mu$m were obtained simultaneously in the SPIRE large photometric scanning 
observing mode (``SpirePhotoLargeScan'') on 29 June 2010 over one pass lasting 
48.4 minutes (ObsID: 1342199389).  The SPIRE scan speed was 30$^{\prime\prime}$~s$^{-1}$ 
(the medium scan rate).  
Table~\ref{tab:obs} summarizes the {\it Herschel} observations.

The PACS and SPIRE data were reduced separately following standard procedures 
in HIPE version~13 \citep{Ott2010} using calibration set~65.  
Table~\ref{tab:fluxes} lists the measured fluxes or upper limits for each band.
PACS aperture photometry is measured using 12$^{\prime\prime}$ and 22$^{\prime\prime}$
apertures for the 70 and 160~$\mu$m bands, respectively.
Uncertainties for the PACS photometry values are computed using several apertures on
the background.
All uncertainties are 1~$\sigma$.
Table~\ref{tab:fluxes} also lists the beam sizes for each band \citep{Vandenbusscheetal2010}.

\begin{table}
\caption{{\it Herschel} Measurements of $\tau$ Ceti} 
\label{tab:fluxes}
\begin{tabular}{l|lll}
\hline
$\lambda$ & Flux & Unc. & Beam size \\ 
($\mu$m) & (mJy) & (mJy) & ($^{\prime\prime}$) \\ \hline
70 & 303 & 6 & 5.6 \\ 
160 & 111 & 8 & 11.3 \\ 
250 & 35 & 10 & 18.1 \\ 
350 & $<$28 & - & 25.2 \\ 
500 & $<$20 & - & 36.9 \\ \hline
\end{tabular}
\end{table}

\begin{figure*}
\centering
\includegraphics[scale=0.5]{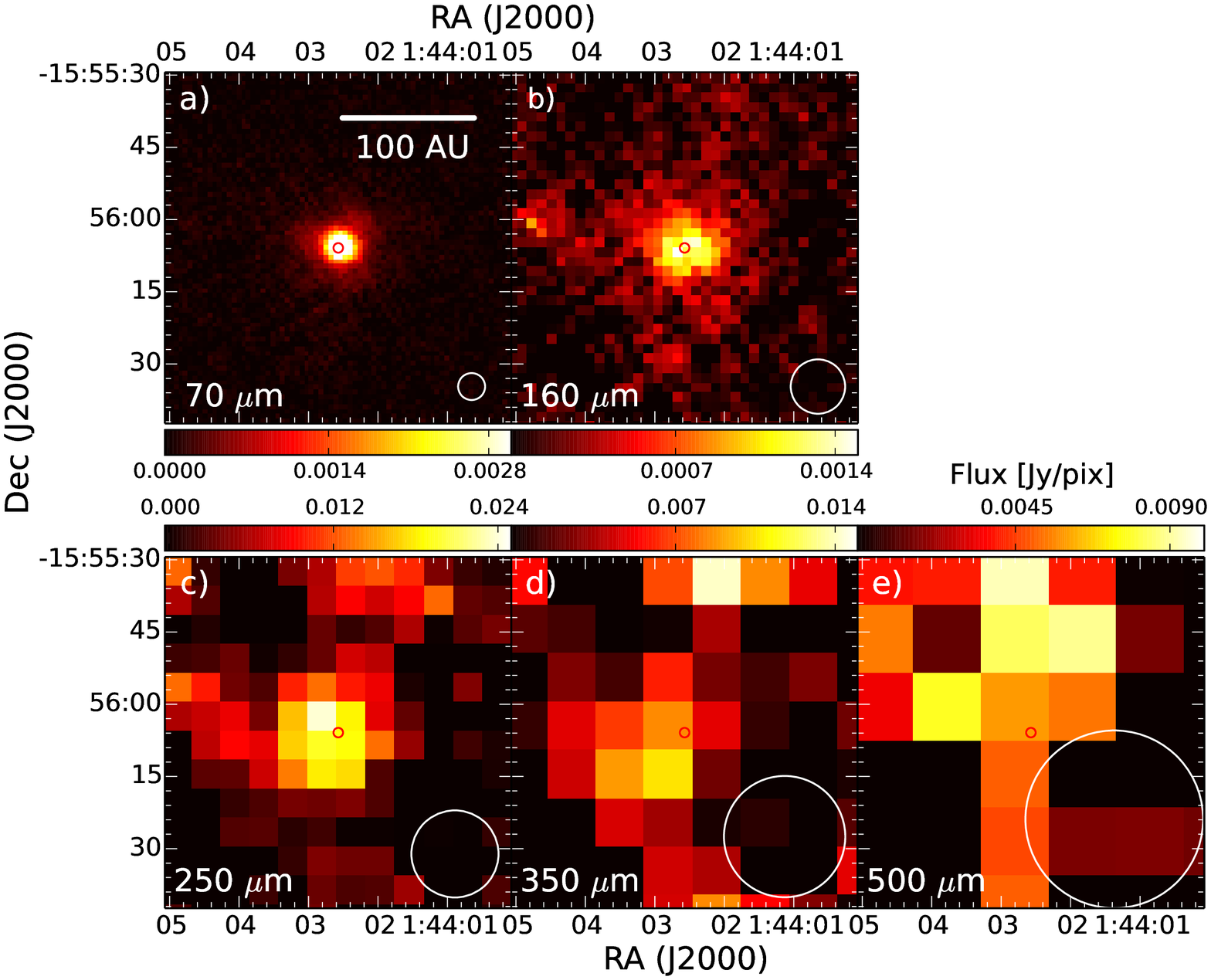}
\caption{
Cropped {\it Herschel} maps of $\tau$~Ceti centered on the position of peak 
70~$\mu$m emission (shown in each panel as a small red circle).  
Beam size at each wavelength is shown for reference in each panel (larger white circles).
A scalebar showing the color range of flux in Jy per pixel is shown above each sub-figure.
a) 70~$\mu$m emission,
b) 160~$\mu$m emission,
c) 250~$\mu$m emission, 
d) 350~$\mu$m emission, and
e) 500~$\mu$m emission.
A bar in a) shows the spatial extent of 100~AU at the distance of $\tau$~Ceti.
}
\label{fig:herschelmaps}
\includegraphics[scale=0.5]{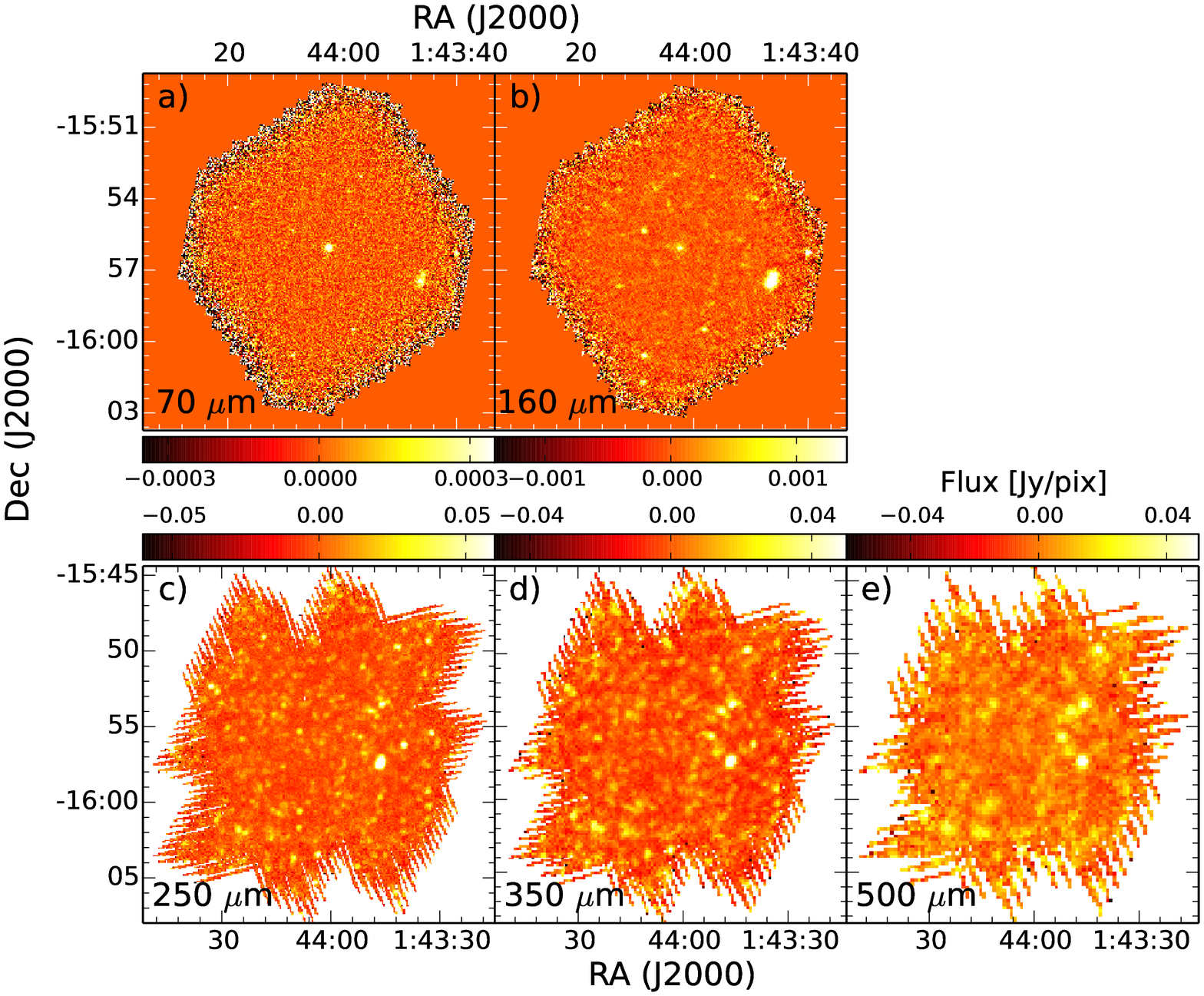}
\caption{
Uncropped {\it Herschel} maps of the $\tau$~Ceti region.  
A scalebar showing the color range of flux in Jy per pixel is shown above each sub-figure
a) 70 $\mu$m emission,
b) 160 $\mu$m emission,
c) 250 $\mu$m emission, 
d) 350 $\mu$m emission, and
e) 500 $\mu$m emission.
}
\label{fig:map}
\end{figure*}

\begin{figure*}
\centering
\includegraphics[scale=0.4]{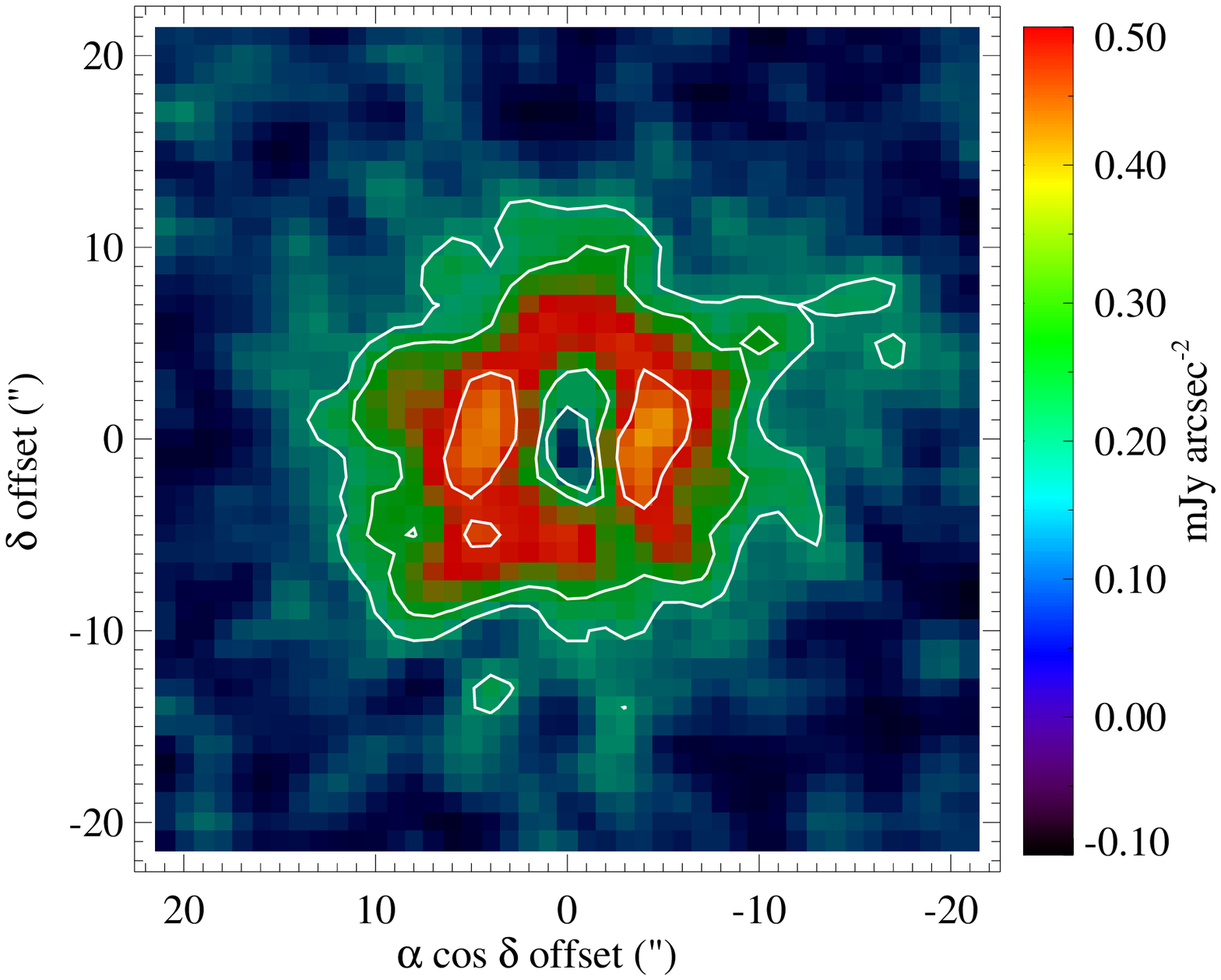} \includegraphics[scale=0.4]{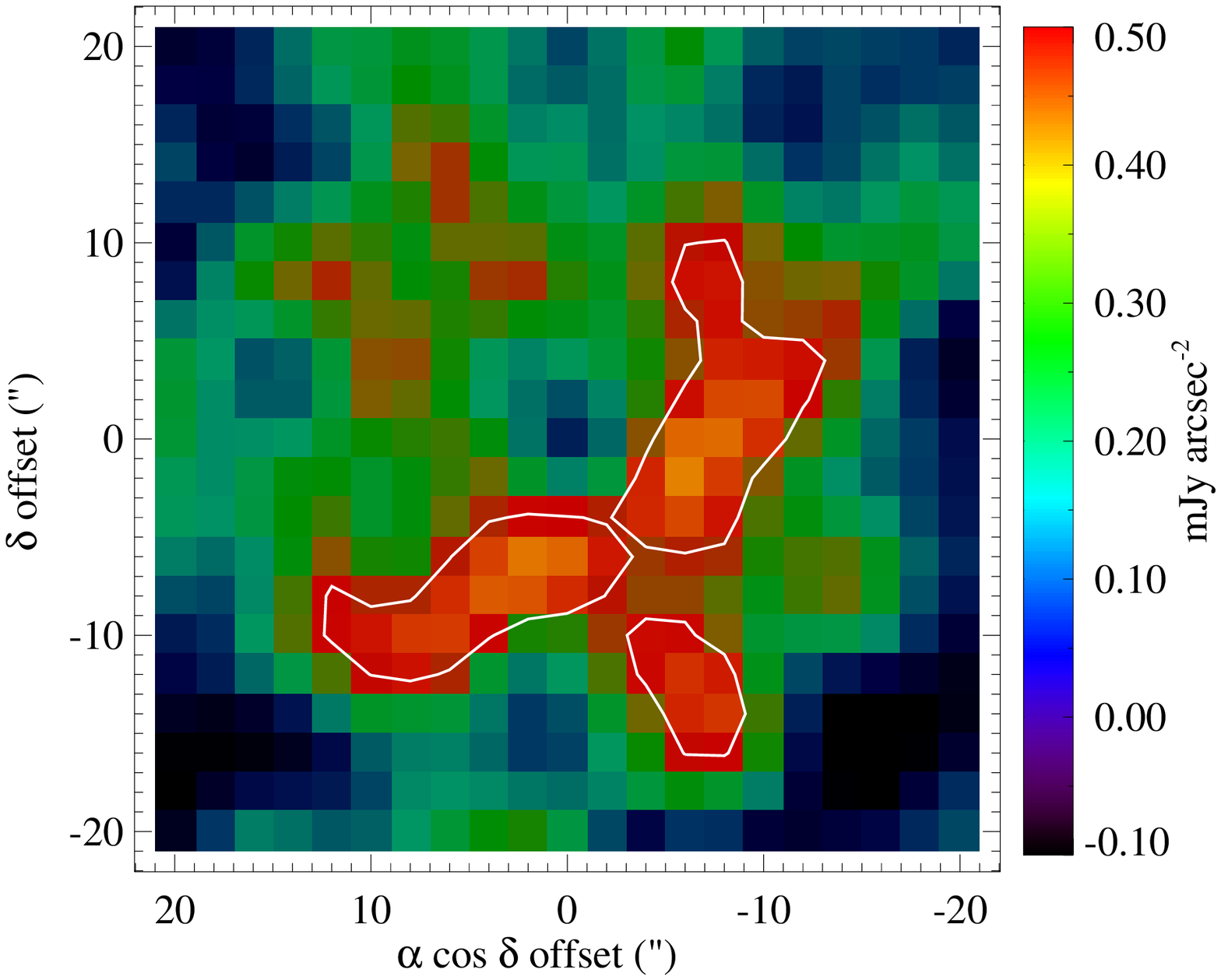}
\caption{
Peak-subtracted images at 70 (left) and 160~$\mu$m (right).
Here a PSF has been scaled to the flux of the peak pixel in each band and subtracted from the images
(see text for details).
Contours show 3~$\sigma$, 5~$\sigma$, and 10~$\sigma$ significance levels of the remaining flux
(only the 3~$\sigma$ contour is visible in the right image).
}
\label{fig:psfsub}
\includegraphics[scale=0.4]{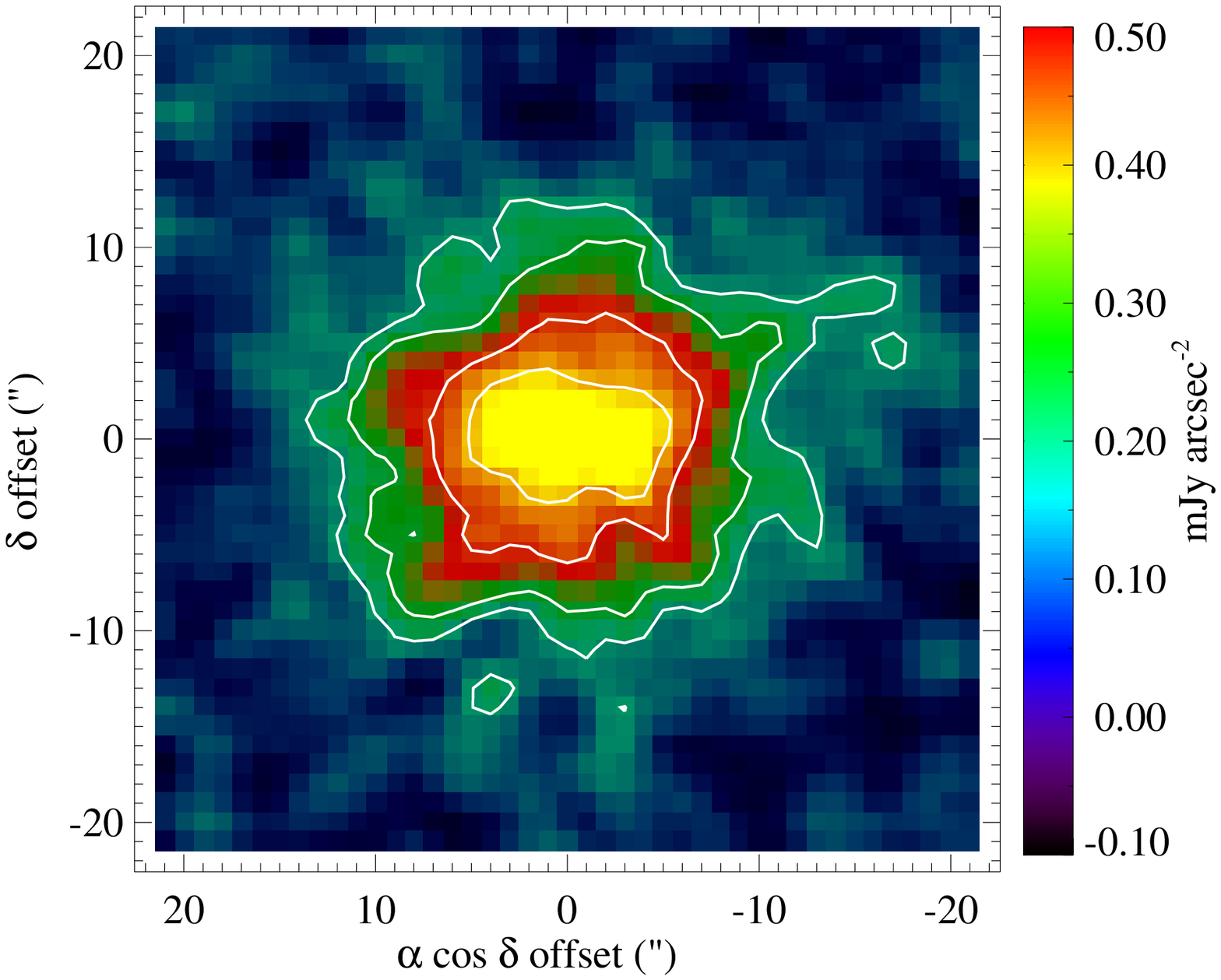} \includegraphics[scale=0.4]{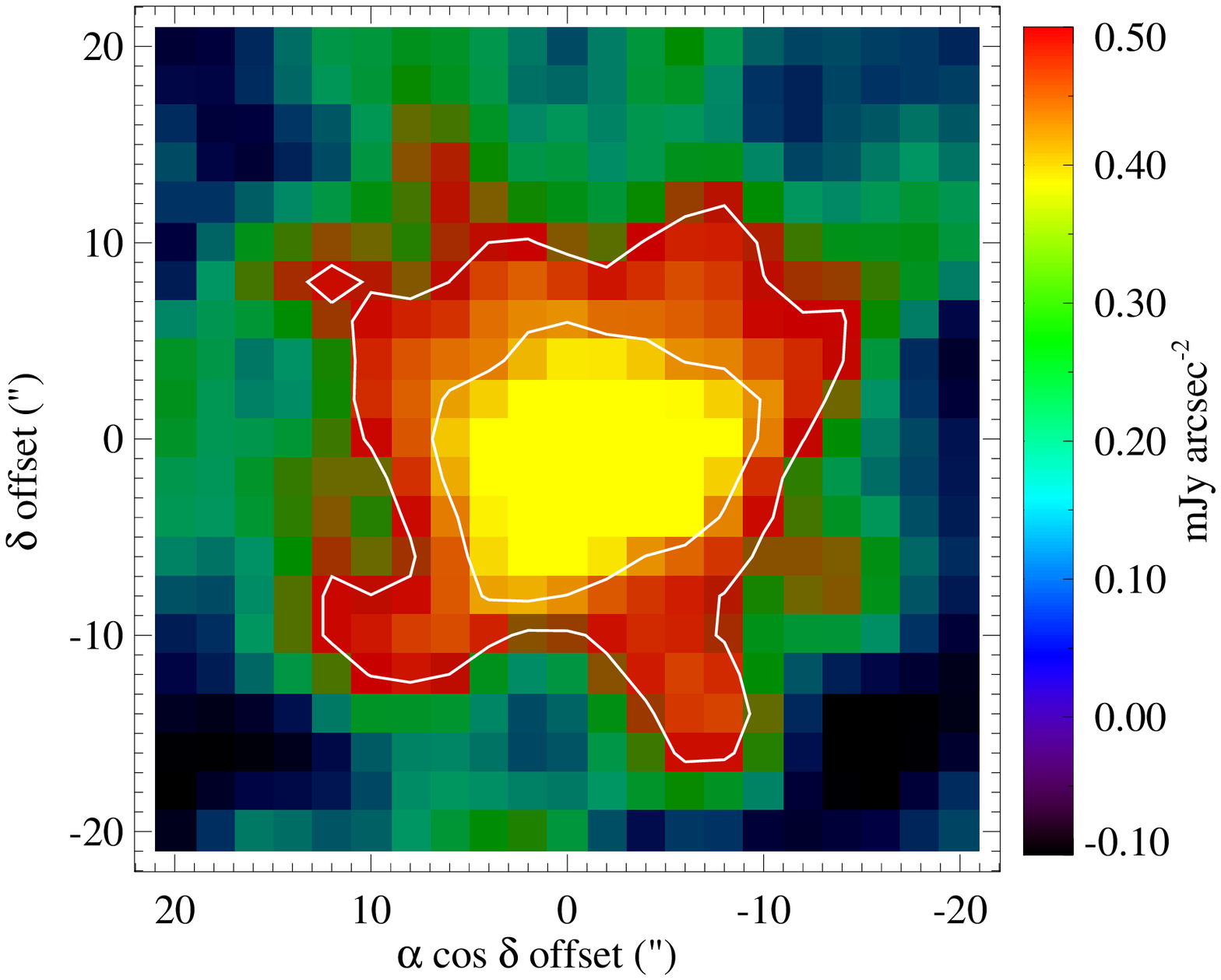}
\caption{
Star-subtracted images at 70 (left) and 160~$\mu$m (right).
Here a PSF has been scaled to the photospheric flux in each band and subtracted from the images.
Contours show 3~$\sigma$, 5~$\sigma$, 10~$\sigma$, and 15~$\sigma$ significance levels of the remaining flux
(only the 3~$\sigma$ and 5~$\sigma$ contours are visible in the right image).
}
\label{fig:starsub}
\end{figure*}

The SPIRE data provide mainly upper limits, as the disk grows fainter and the resolution grows
larger as we proceed to longer wavelengths.
At 250~$\mu$m, assuming the disk is not resolved, the flux comes from the peak pixel, which has 
24~$\pm$~6~mJy, with the uncertainty coming from confusion.
Using the modelling described in Section~\ref{sec:imagemodel}, we find 35~$\pm$~10~mJy.
This is more reliable since the disk may be marginally resolved at 250~$\mu$m.
At 350~$\mu$m there is some emission where $\tau$~Ceti is expected to be, which added to the confusion
limit gives $<$28~mJy.
At 500~$\mu$m, only confusion is visible, giving a limit of $<$20~mJy.

Photometric calibration 
uncertainties are expected to be within 3\% at 70~$\mu$m, 5\% at 160~$\mu$m, 
and 15\% at 250~$\mu$m, 350~$\mu$m, and 500~$\mu$m 
\citep{Poglitschetal2010,Swinyardetal2010,Mulleretal2011}. 

Figure~\ref{fig:herschelmaps} shows a composite of cropped {\it Herschel} 
maps at 70~$\mu$m, 160~$\mu$m, 
250~$\mu$m, 350~$\mu$m, and~500 $\mu$m, centered at a position of 01:44:02.6, 
-15:56:05.8 (J2000), the position of peak emission at 70~$\mu$m.  This position
is within $\sim$1$^{\prime\prime}$ of the expected position of $\tau$~Ceti, given 
both its ICRS coordinates (i.e., 01:44:04.08,-15:56:14.9) and proper motion 
($\mu_{\alpha}$ = -1.72105 $^{\prime\prime}$ yr$^{-1}$, $\mu_{\delta}$ = 
$+$0.85416 $^{\prime\prime}$ yr$^{-1}$) from the Hipparcos Catalogue 
\citep{vanLeeuwen2007}.

$\tau$~Ceti at 70 $\mu$m in Figure~\ref{fig:herschelmaps}a 
is bright and quite compact; the three 
``lobes" located in the NNW, SSW, and ENE directions are artifacts 
of the PACS beam at 70~$\mu$m.  
Given these secondary beam features, it is
not possible to tell by eye if any 70~$\mu$m emission is extended.  
The source at 160~$\mu$m in Figure~\ref{fig:herschelmaps}b 
is less bright but still compact.  
At 250~$\mu$m (see Figure~\ref{fig:herschelmaps}c), 
the source is less significant but the star-centered emission is associated with 
$\tau$~Ceti.  
At 350~$\mu$m and 500~$\mu$m 
(Figures~\ref{fig:herschelmaps}d and \ref{fig:herschelmaps}e respectively), any emission at 
the $\tau$~Ceti position is hard to distinguish from the background confusion.  

Figure~\ref{fig:map} shows the uncropped {\it Herschel} maps at each wavelength, stretched in 
colour to emphasize faint background sources.  At 70 $\mu$m (Figure~\ref{fig:map}a), $\tau$~Ceti
dominates the image but a few other background sources are seen.  Most notably,
the extended galaxy MCG-03-05-018 located to the WSW at 01:43:46.8,-15:57:29 
(J2000) is also detected.  Moving to 160 $\mu$m (Figure~\ref{fig:map}b), $\tau$~Ceti is no 
longer the brightest object seen; MCG-03-05-018 and many other background 
objects are brighter.  Also, more faint background sources are seen.  At 250 
$\mu$m, 350 $\mu$m, and 500 $\mu$m (Figures~\ref{fig:map}c, d, and e, respectively), 
though some emission may be associated 
with $\tau$~Ceti (particularly at 250 $\mu$m), it is barely distinguishable from 
emission of background objects.  Note that MCG-03-05-018 is clearly detected
in all five {\it Herschel} bands.

\section{$\tau$ Ceti's Debris Disk} \label{sec:disk}

It is not obvious from Figure~\ref{fig:herschelmaps} that a disk is present due to large 
contrast with the star. 
Therefore, we present peak- and star-subtracted images to highlight the extended 
disk structure.
Figure~\ref{fig:psfsub} shows the images of the flux toward 
$\tau$~Ceti at 70 and 160~$\mu$m, where the point-spread function (PSF) has been scaled to the
value of the peak pixel and subtracted.
This makes it clear that there is extended structure around $\tau$~Ceti,
visible at both 70 and 160~$\mu$m.
For comparison, Figure~\ref{fig:starsub} shows the star-subtracted images at the same wavelengths,
where the PSF has been scaled to $\tau$~Ceti's photosphere and subtracted;
contours give significance of the remaining flux.

\begin{figure}
\centering
\includegraphics[scale=0.4]{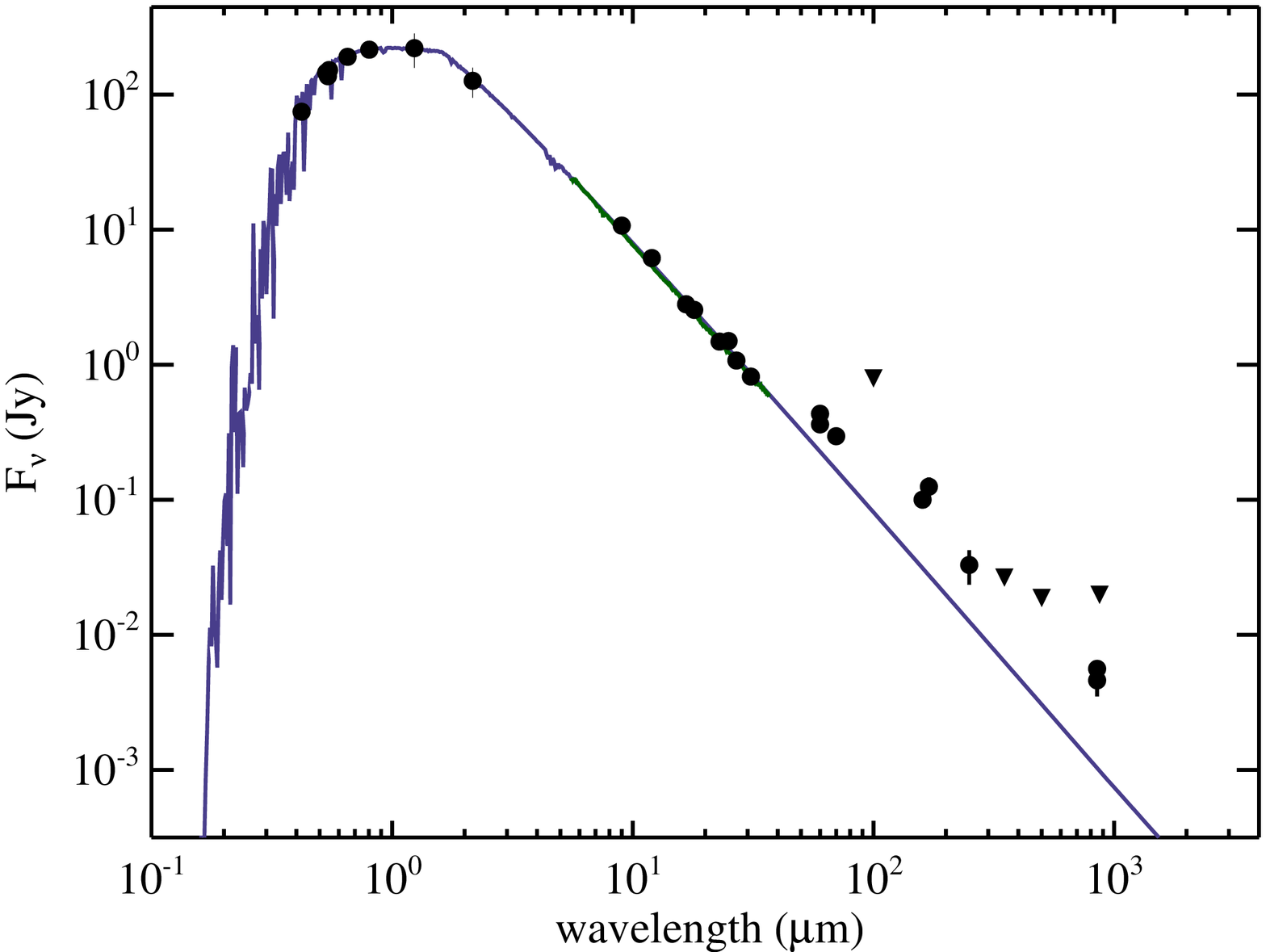}
\caption{
Observed SED of $\tau$~Ceti.  
Circles are measured fluxes, while triangles show upper limits. 
{\it Spitzer} IRS data are shown in green, on top of the photospheric model (blue line).
The observed data include fluxes obtained from optical 
\citep{Mermilliod2006,Hogetal2000,HauckMermilliod1998,Perrymanetal1997,Bessel1990}, 
2MASS \citep{2MASS}, {\it Spitzer} \citep{Chenetal2006}, AKARI \citep{Ishiharaetal2010}, 
IRAS \citep{Moshiretal1990}, ISO \citep{Habingetal2001}, 
JCMT \citep[][Greaves et al.\ in prep.]{Greavesetal2004}, and HHT \citep{Holmesetal2003dd},
as well as the {\it Herschel} PACS and SPIRE values and upper limits.
See Tables~\ref{tab:fluxes} and \ref{tab:allfluxes} for values.
}
\label{fig:SED}
\end{figure}

Figure~\ref{fig:SED} shows the observed flux density distribution 
(hereafter referred to as spectral energy distribution; SED) of $\tau$~Ceti,
using data obtained from the literature and {\it Herschel} (see caption and Table~\ref{tab:allfluxes} 
for specific references).  
Figure~\ref{fig:SED} also shows a stellar photosphere model fit 
to the data.  
As can be easily seen, the stellar model fits the data from 
optical to mid-infrared wavelengths (24~$\mu$m).  
At longer (PACS) wavelengths,
however, the observed fluxes are significantly higher than those expected 
from the photosphere alone.  
For example, the observed 160~$\mu$m flux is 111~$\pm$~8~mJy  
while the expected 160~$\mu$m flux from the photosphere 
is 31.1~$\pm$~0.4~mJy.  

\begin{table*}
\caption{Observational Data from the Literature}
\label{tab:allfluxes}
\begin{tabular}{lllll}
\hline
Band 	&	$\lambda$	&	Obs.\ Flux	&	Uncertainty	&	Citation	\\
	&	($\mu$m)	&	(Jy)	&	(Jy)	&		 \\ \hline
$U_{\rm J}$	&	0.364	&	146.6	&	3.2	&	\citet{Mermilliod2006}	\\
$B_{\rm T}$	&	0.42	&	74.7	&	1.1	&	\citet{Hogetal2000}	\\
$B_{\rm J}$	&	0.442	&	146.4	&	3.2	&	\citet{Mermilliod2006}	\\
$V_{\rm T}$	&	0.532	&	145.7	&	1.4	&	\citet{Hogetal2000}	\\
$H_{\rm p}$	&	0.541	&	136.7	&	0.68	&	\citet{Perrymanetal1997}	\\
$V_{\rm J}$	&	0.547	&	151.9	&	3.2	&	\citet{Mermilliod2006}	\\
$R_{\rm C}$	&	0.653	&	190.2	&	3.5	&	\citet{Bessel1990}	\\
$I_{\rm C}$	&	0.803	&	214.9	&	3.9	&	\citet{Bessel1990}	\\
$J$	&	1.24	&	220.6	&	63	&	\citet{Cutrietal2003}	\\
$K_{\rm s}$	&	2.16	&	126.4	&	32	&	\citet{Cutrietal2003}	\\
IRS1$^a$ &	6.5	&	18.15	&	0.65	&	\citet{Chenetal2006}	\\
IRS2$^a$ &	8.69	&	10.19	&	0.24	&	\citet{Chenetal2006}	\\
AKARI9	&	9	&	10.71	&	0.17	&	\citet{Ishiharaetal2010}	\\
IRS3$^a$	 &	11.4	&	5.885	&	0.13	&	\citet{Chenetal2006}	\\
IRAS12	&	12	&	6.158	&	0.42	&	\citet{Moshiretal1990}	\\
IRS4$^a$	 &	16.6	&	2.801	&	0.073	&	\citet{Chenetal2006}	\\
AKARI18	&	18	&	2.544	&	0.071	&	\citet{Ishiharaetal2010}	\\
IRS5$^a$ &	22.9	&	1.509	&	0.039	&	\citet{Chenetal2006}	\\
IRAS25	&	25	&	1.503	&	0.14	&	\citet{Moshiretal1990}	\\
IRS6$^a$ &	27	&	1.094	&	0.025	&	\citet{Chenetal2006}	\\
IRS7$^a$ &	31	&	0.8393	&	0.021	&	\citet{Chenetal2006}	\\
ISO60	&	60	&	0.433	&	0.037	&	\citet{Habingetal2001}	\\
IRAS60	&	60	&	0.3978	&	0.048	&	\citet{Moshiretal1990}	\\
IRAS100	&	100	&	0.8253	&	0.27	&	\citet{Moshiretal1990}	\\
ISO170	&	170	&	0.125	&	0.021	&	\citet{Habingetal2001}	\\
SCUBA	&	850	&	0.0058	&	0.0006	&	\citet{Greavesetal2004}	\\
SCUBA-2	&	850	&	0.005	&	0.001	&	Greaves et al.\ (in prep.)	\\
HHT870	&	870	&	$<$0.0198	&	-	&	\citet{Holmesetal2003dd}	\\ \hline
\multicolumn{5}{l}{$^a$ IRS values are binned from spectral data in \citet{Chenetal2006}.}
\end{tabular}
\end{table*}

The appearance of excess emission in the SED suggests that $\tau$~Ceti is indeed 
surrounded by cooler dust.  The narrow wavelength range of the detected excess 
and the lack of widely extended PACS emission, however, put constraints on the 
location of this dust.  
On one hand, the fact that the excess is seen only at
wavelengths longer than 24~$\mu$m suggests the dust is cold and thus situated 
at relatively large distances from the star.  
On the other hand, the small scale of the emission in the PACS images suggests the dust 
also cannot be too far from the star.  
In the following, we describe a simple debris disk model that remains consistent 
with the observed optical to mid-infrared emission, and that also can reproduce well the fluxes and 
extents of the observed far-infrared (PACS) emission.

\subsection{Modelling the Disk}

We modelled the $\tau$~Ceti disk images and spectrum in a two-step process. 
The images provide all-important constraints on the spatial structure, 
which we obtain first by reproducing the PACS and SPIRE 250~$\mu$m
images using a simple dust disk model (Section~\ref{sec:imagemodel}). 
This process does not yield definitive results because the PACS image resolutions limit 
what can be inferred about the disk inner edge. 
Using the results from the spatial modelling and an additional assumption 
of specific grain properties to model the disk spectrum (Section~\ref{sec:sedmodel})
yields further constraints on the disk inner edge.

\subsubsection{Image-Based Disk Model} \label{sec:imagemodel}

\begin{figure*}
\centering
\includegraphics[scale=0.65]{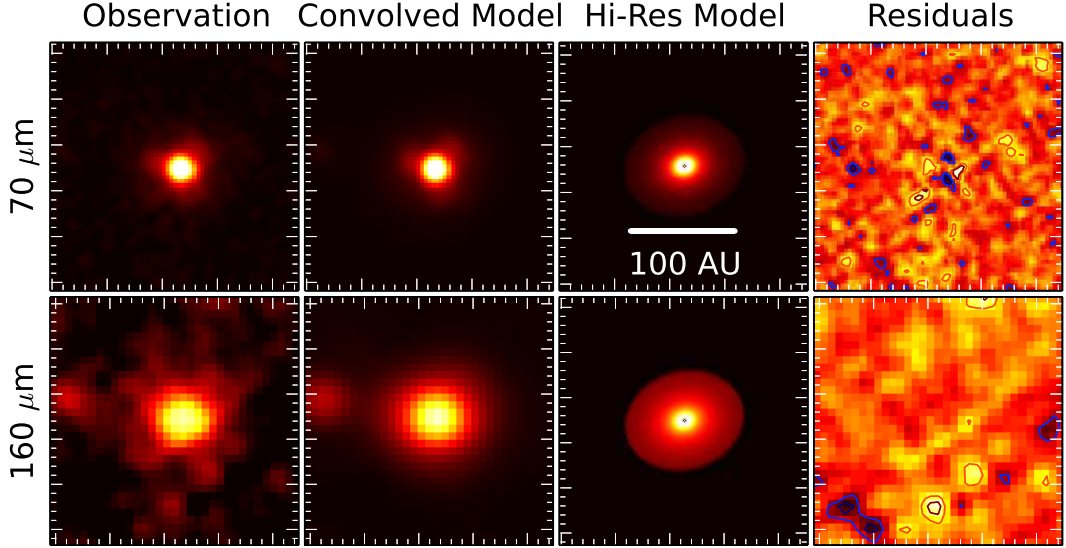}
\caption{
An example of a disk model
that matches the {\it Herschel} images.  Panels are on the same scale as Figure~\ref{fig:herschelmaps},
and a 100~AU scale bar is shown for reference.
Top panels show the {\it Herschel} data and model
at 70~$\mu$m, lower panels at 160~$\mu$m.  From left to
right, panels show the {\it Herschel} data, the model convolved to the same resolution as the
data, a high resolution version of the model, and the residuals of data-model (-3~$\sigma$, -2~$\sigma$, 
2~$\sigma$, and 3~$\sigma$ contours).  
Some 3~$\sigma$ residuals are still visible in the 70~$\mu$m
image; these are probably due to imperfect fitting of the beam (see text).
}
\label{fig:modeldisk}
\end{figure*}

We first use the PACS and SPIRE 250~$\mu$m images to fit a physical model for the disk
structure. The 250~$\mu$m image has little spatial information, but 
we can measure the model flux as a check on the above photometry 
(finding 35~$\pm$~10~mJy, larger than the 24~mJy estimated above if the disk were unresolved at 250~$\mu$m).
Our spatial model has been used previously to model {\it Herschel}-resolved debris
disks \citep[e.g.][]{Kennedyetal2012,Kennedyetal2013}, 
and generates a high-resolution image of an
azimuthally symmetric dust distribution with a small opening angle, as viewed from a
specific direction. These models are then convolved with a point spread function model
(observations of calibration star $\alpha$~Tau) for comparison with the observed
disk (Figure~\ref{fig:modeldisk}). 
The best-fitting model is found using by-eye approximation followed by
least-squares minimization. As the entire multi-dimensional parameter space was not
searched, the model presented is not necessarily unique, but provides a good indication
of the probable disk structure.
By checking the fit of different parameter combinations, we were able to get a good feel for
how well-constrained the different parameters are and feel we have converged on a good model,
within observational errors.

Due to the limited resolution of the images of $\tau$~ Ceti, our model disk is a simple
power-law in radial surface density ($\Sigma \propto r^\gamma$), which extends from
$r_{\rm in}$ to $r_{\rm out}$. This approach allows us to test whether or not the disk is
radially extended. 
We use the simple assumption of a blackbody temperature law ($T =
T_{\rm 1AU} r^{-0.5}$, with $T_{\rm 1AU}$ being the disk temperature at 1~AU and radius $r$ in AU). 
Given that the disk temperature and surface
density are degenerate without multiple well-resolved images, neither $\gamma$ 
nor $T_{\rm 1AU}$ is well constrained. Physically,
$T_{\rm 1AU}$ should be greater than about 230~K, because this is the temperature that
grains with blackbody absorption and emission would have at 1~AU from $\tau$~Ceti. 
Temperatures up to factors of 3-4 greater are possible if the emission comes
primarily from small grains, which emit inefficiently at long wavelengths and have higher
temperatures to maintain energy equilibrium \citep[e.g.][]{Boothetal2013}.

The model also includes a background source to the east that is only visible in the 160~$\mu$m image.
This does not significantly affect any of the parameters of the fit,
but does allow us to better estimate the 250~$\mu$m disk flux.

To model the {\it Herschel} images of $\tau$~Ceti, we initially tried a simple narrow ring. 
We found that this model failed to reproduce the observed images and conclude that the emission
is radially extended. We therefore allowed $r_{\rm in}$ and $r_{\rm out}$, and $\gamma$, the
surface density power-law exponent, to vary independently. 
The low surface brightness of the disk in the {\it Herschel} images limits our ability to constrain 
the disk model parameters.
Primarily, a degeneracy between disk surface density profile, the inner edge location, and the dust 
temperature allowed a range of models to reproduce the data. 
For a $\gamma=-1$ model, the disk is centrally concentrated and the best fitting inner edge is at 
about 10~AU with $T_{\rm 1AU}\sim 380$~K. 
For a flat profile ($\gamma=0$) the disk is less centrally concentrated and the inner edge is closer, 
around 2~AU (and $T_{\rm 1AU}$ is the same). 
For a radially increasing surface density ($\gamma=1$), the inner edge is around 3~AU and $T_{\rm 1AU}\sim 180$~K.
We return to this issue
when considering SED models that make assumptions about grain properties in 
section~\ref{sec:sedmodel} below.

The low surface brightness of the disk in the {\it Herschel} images limits our ability to constrain 
the disk model parameters.
However, despite the degeneracies between $T_{\rm 1AU}$, $r_{\rm in}$ and $\gamma$, 
the best fitting models have similar inner radii of 2-3~AU 
(with large uncertainty, acceptable fits range from roughly 1-10~AU), and
outer radii of 55~AU~$\pm$~5~AU. 
The disk may however extend to larger radii at a level undetectable by these observations.
The disk geometry is constant across different models,
with a disk inclination (i.e., from face-on) of 35$^\circ$ and position angle (East of
North) of 105$^\circ$. Using brute-force grid calculations we estimate that the 1$\sigma$
uncertainty in these angles is about 10$^\circ$. 

Figure~\ref{fig:modeldisk} shows an example of a well fitting model 
with $\gamma=0$. 
Some residual structure is seen near the star at
70~$\mu$m, and very similar structure is seen for different models.
We suspect it arises
due to the brightness of $\tau$ Ceti itself; the high signal-to-noise ratio of the
stellar emission means that the PSF model used ($\alpha$ Tau) must be a very good match
to the PSF for the $\tau$ Ceti observation. \citet{Kennedyetal2012} showed that the PACS
70~$\mu$m PSF varies at the 10\% level, which is a probable reason for the non-zero
residuals near the star.

\subsubsection{SED-Based Disk Model} \label{sec:sedmodel}

\begin{figure}
\centering
\includegraphics[scale=0.5]{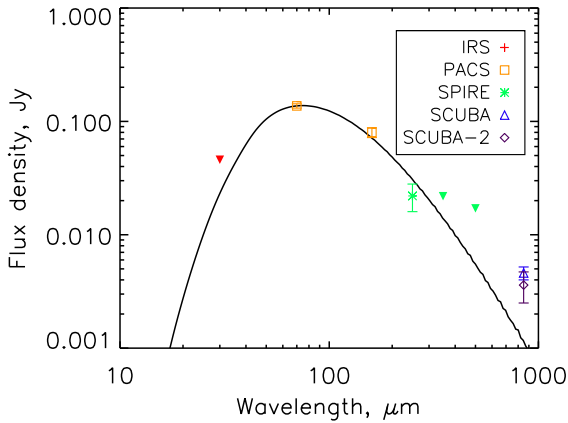}
\caption{
Fitting the SED with a realistic grain model, using a flat surface density profile ($\gamma=0$).
The {\it Herschel} PACS and SPIRE 250~$\mu$m datapoints are used in the fit, along with
upper limits from {\it Spitzer} IRS and {\it Herschel} SPIRE 350~$\mu$m and 500~$\mu$m.
SCUBA and SCUBA-2 850~$\mu$m data are also shown, but are not used in the fit.
See legend for symbols.
}
\label{fig:SEDfit}
\end{figure}

Though in general the shape of a modified blackbody provides a good approximation to the 
emission profile of dust grains, it cannot tell us much about the properties of the grains. 
A better approximation of the dust grains can be found by taking into account their 
optical properties and the size distribution of dust in the disk. 
Accordingly, we follow the model described in \citet{WyattDent2002} and similarly assume non-porous,
amorphous silicate grains with an organic refractory mantle. 
The silicate core makes up 1/3 of the grains, which have an amorphous carbon coating.  
Grains are assumed to be spherical, and absorption efficiencies are calculated
using Mie theory, Rayleigh-Gans theory, or geometric optics in the appropriate limits
\citep[see also][]{LiGreenberg1997}. 
We assume these grains follow a size distribution power law of 
$n(D)\rm{d}D\propto D^{-3.5}\rm{d}D$ 
from a minimum grain size, $D_{\rm min}$, 
to an arbitrarily large grain size \citep{Dohnanyi1969}. 
Many of the properties of the grain model are degenerate. 
For instance, a 
shallower size distribution has the same effect as increasing the minimum grain size, 
and the exact composition of the grains cannot easily be determined without spectral features. 
For this reason, we only varied $D_{\rm min}$ and the inner edge of the
disk (since this was poorly constrained from the image-based modelling). 
We also tested two values of the surface density power law, $\gamma$
equal to 0 and 1 since this quantity is also poorly constrained from the image-based modelling.

No SED model can be found that fits both the SCUBA and PACS 160~$\mu$m photometry, suggesting 
that a separate, cooler disk component may be required to fully explain the sub-mm observations,
since a slope that fits both these points is too shallow to even be fit by a perfect blackbody.
Further investigation of the multi-epoch SCUBA and SCUBA-2 data is left to Greaves et~al.\ (in prep).

For the following we therefore focus on fitting the model to the PACS data, the SPIRE data 
and the upper limit from the Spitzer IRS spectral data. 
We find the best fitting model to have a flat surface density profile ($\gamma=0$), 
a minimum grain diameter of $15\pm8$~$\mu$m and an inner radius of 6$^{+15}_{-4}$~AU 
(parameter uncertainties calculated using a $\chi^2$ cut). 
The best fit is shown in Figure~\ref{fig:SEDfit}. 
There is some anti-correlation between the minimum diameter and inner radius such that 
models with a larger inner radius require a smaller minimum grain size. 
Models with a rising surface density of $\gamma=1$ can still plausibly fit the photometry 
with a minimum grain size of 8~$\mu$m and an inner radius between 1~AU and 17~AU, 
although this provides a poorer fit to the data.

Unfortunately, these SED models were not very sensitive to different disk inner edges, and the results 
of this modeling technique, while agreeing with the results of the image-based model, did not
provide any stronger constraints on the inner edge of the disk.

\subsubsection{Disk Properties Inferred from Both Models}

To summarize the findings of both models, 
$\tau$~Ceti's disk extends from a radius similar to the inner Solar System ($\sim$1-10~AU) to 
just outside the distances inhabited by the main classical Kuiper Belt.

The image-based model's uncertainty in the inner edge locations arises from the variations
in the 70~$\mu$m beam shape and the resolution limits in both PACS wavelengths. 
The uncertainty in the inner edge as predicted by the SED-based model can be
attributed to the calibration uncertainties in the {\it Spitzer} IRS spectrum
and a small number of photometric measurements of the excess,
while the outer edge uncertainty is mainly due to the low surface brightness in the
PACS 160~$\mu$m image.

We do not give a mass estimate for our dust models, as the uncertainties due to
assumptions about the grains give disk masses that vary by orders of magnitude.
The most accurate disk masses come from submillimeter fluxes, temperatures, and 
opacities, thus we leave calculation of the disk mass in the $\tau$~Ceti system
to the forthcoming SCUBA and SCUBA-2 analysis (Greaves et~al.\ in prep.)

In the next section, we will use the (limited) constraints imposed by 
the disk to investigate the validity of a proposed planetary system.

\section{$\tau$ Ceti's Possible Planetary System} \label{sec:planets}

Though previous studies failed to find planets around $\tau$~Ceti 
using the radial velocity technique \citep[e.g.][]{Pepeetal2011},
\citet{Tuomietal2013} report evidence for a five planet system after extensive
modelling and 
Bayesian statistical analysis using combined radial velocity data from three
different planet surveys.  

The most likely system found by Tuomi et~al.\
consists of five super-Earths, ranging in mass ($M \sin i$) from 
2.0-6.6 $\Mearth$, with small-to-moderate eccentricities ($\sim$0-0.2), in a tightly-packed
configuration with semimajor axes ranging from 0.105-1.35~AU.
\citet{Tuomietal2013} show that their system is stable based on Lagrange stability
thresholds, but do not perform detailed numerical integrations.

We note that the periodic RV signals detected by \citet{Tuomietal2013}
were only interpreted as planets by these authors with caution; it is possible that
the signals are from another source, such as stellar activity or instrumental bias,
although there is no direct evidence in favour of these alternative interpretations either.
In this section we investigate the stability of the proposed planet system, and 
assuming that the planetary system is real, use it to place constraints
on the inner disk edge using dynamical simulations.

\subsection{System Inclination}

Since the planets' existence was surmised using RV data, we have no direct 
information on the inclination of the planetary system.
In addition to the coplanar precedent of our own Solar System,
several recent studies find evidence that star-planet-disk systems 
without hot Jupiters should be well-aligned.
For example,
\citet{Kennedyetal2013} discuss the HD~82943 system, where the star, planets, and debris 
disk have well-measured inclinations, and are all coplanar within $\sim$10$^{\circ}$.
Furthermore, \citet{Greavesetal2014} find that, out of 11 systems with {\it Herschel}-resolved disks 
and well-measured stellar inclinations, all are consistent with being coplanar.
\citet{Watsonetal2011} measure the rotational axes of stars with resolved debris disks,
and reach the same conclusion.
Studies of {\it Kepler}-discovered multiplanet systems 
\citep{SanchisOjedaetal2012,Hiranoetal2012,Albrechtetal2013,Chaplinetal2013}
also have found that the orbital planes of the planetary systems tend to be well-aligned 
with the equators of the host stars.
We believe these studies provide ample evidence that compact, low-mass planetary systems 
like $\tau$~Ceti are usually well-aligned systems.

\begin{table}
\caption{Properties of a Stable Planet System around $\tau$ Ceti} 
\label{tab:planets}
\begin{tabular}{l|lll}
\hline
planet & $M^a$ & $a$ & $e$ \\ 
  & ($\Mearth$) & (AU) &  \\ \hline
a & 4.0 & 0.105 & 0.16 \\ 
b & 6.2 & 0.195 & 0.03 \\ 
c & 7.2 & 0.374 & 0.08 \\ 
d & 8.6 & 0.552 & 0.05 \\ 
e & 13.2 & 1.35 & 0.03 \\ \hline
\multicolumn{4}{l}{$^a$ Planet masses are given assuming $i_{\rm sys}=30^{\circ}$}
\end{tabular}
\end{table}

\citet{Greavesetal2004} used the low rotational velocity measured by \citet{SaarOsten1997}
as evidence that we are viewing $\tau$~Ceti within 40$^{\circ}$ of pole-on, which was
inconsistent with their measurements of the disk inclination.
However, at the SCUBA resolution and wavelength, background confusion made it difficult to 
measure the disk inclination.
The analysis of the {\it Herschel} images presented here has made it clear that the 
disk is close to face-on, consistent with being aligned with the equatorial plane of $\tau$~Ceti.

Assuming the best-fitting inclination for the disk ($\sim$30$^{\circ}$)
equals the inclination for $\tau$~Ceti and its planetary system,
the best-fit values of $M \sin i$ given in \citet{Tuomietal2013} should be doubled.
As found in Section~\ref{sec:planetsystem} below, 
such masses still allow a dynamically stable configuration for the planets.
Table~\ref{tab:planets} gives the masses, semimajor axes, and eccentricities for 
this system.
We note that, however, this is only one possible configuration of planets that satisfies
both the requirement of long-term stability and the 
Bayesian analysis of \citet{Tuomietal2013}.

\subsection{Dynamical Simulations}

We perform numerical simulations using {\tt swift-rmvs4} \citep{LevisonDuncan1994}, 
with a timestep of 0.002 years (0.73 days). 
This allows $>$15 timesteps
per orbit for accurate calculation of the positions of all five planets, including the 
innermost planet, with an orbital period of only 14 days.
All of our simulations were carried out on the Canadian Advanced Network for Astronomical
Research \citep[CANFAR;][]{Gaudetetal2009}.

For our dynamical simulations, we ignore the mass of the disk  
since the largest reasonable estimate of $\tau$~Ceti's disk mass is about 10\% of the mass of the
outermost planet \citep[$M_{\rm disk}\simeq1\Mearth$;][]{Greavesetal2004}.
While \citet{MooreQuillen2013} find that a disk mass this high relative to the planet masses can affect the dynamical stability lifetime of a planetary system, the system they modelled (HR~8799)
extends to much larger separations from the star than that of $\tau$~Ceti.  
HR~8799b, the outermost planet in the system, has a semimajor axis of 68~AU \citep{Maroisetal2008}
and the planetesimal disk extends from 100-310~AU \citep{Matthewsetal2014}.
If the HR~8799 system is scaled down so that the orbit of HR~8799b matches the outermost planet in 
the $\tau$~Ceti system, HR~8799's entire debris disk would extend to only $\sim$3-6~AU,
while in reality, $\tau$~Ceti's disk mass is actually spread out to $\sim$55~AU.
Given the small semimajor axes of all the planets in the $\tau$~Ceti system, we therefore believe
the contribution of the disk mass to the stability of the system is negligible.

We performed two types of simulations: planetary system stability and disk
orbit stability.
Planet stability simulations were run for 100 Myr, 
corresponding to over two billion orbits of the innermost planet,
while disk simulations were run for 10 Myr with many massless test particles included in addition
to the five planets to diagnose stable orbits for small bodies.
Planetary systems were deemed unstable if any planet's semimajor axis changes by 
$>1\%$ over the course of an integration.
The same change in semimajor axis is used to diagnose stable versus unstable 
disk particle orbits.

Given the infinite range of possible starting conditions for this multiplanet system,
we chose a few representative possibilities and investigated the stability of those 
before proceeding to disk simulations.
We found the highest eccentricities allowed by the statistical analysis of \citet{Tuomietal2013}
yielded unstable planetary systems.
The more moderate (best-fit) eccentricities and very low eccentricities result in
planetary systems stable on timescales of 100~Myr, even when planet masses are increased
by a factor of 1/$\sin i_{\rm sys}$, up to inclinations out of the sky plane as low as 5$^{\circ}$,
nearly perpendicular to our line-of-sight.

\subsubsection{Disk Simulations} \label{sec:planetsystem}

After confirming that the planetary system is stable over a reasonable range of possible orbital
configurations, we use one stable planetary system (Table~\ref{tab:planets}) as part of
another set of simulations.
Here we quantify the stability of small body orbits near 
planets in order to find where the debris disk would be stable over long timescales.
The small bodies are represented by massless test particles in these simulations.

\begin{figure}
\centering
\includegraphics[scale=0.6]{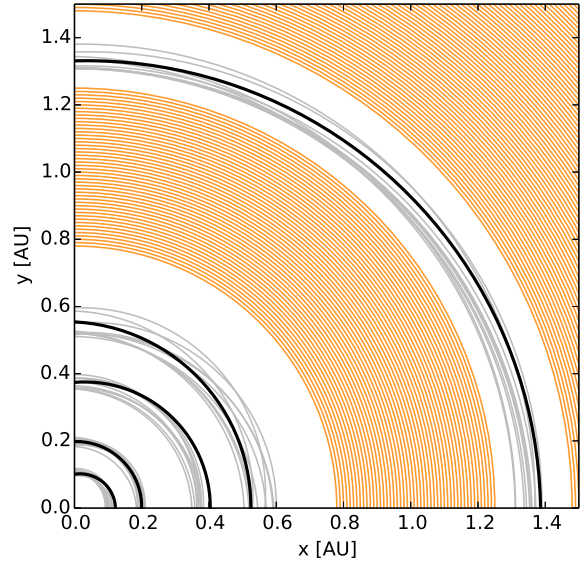}
\caption{
Initial (coplanar) orbits of 5 planets are shown in black, using masses
and orbital elements as given in Table~\ref{tab:planets}.
Orbital elements which were unconstrained by the \citet{Tuomietal2013}
analysis ($\Omega$, $\omega$, and $\mathcal{M}$) are chosen at random.
Gray lines show how the orbits evolve over the course of a 100~Myr integration.
Stable disk particle orbits (surviving a 10~Myr integration) are shown in orange.
}
\label{fig:diskorbits5pl}
\end{figure}

The fairly small planets of the $\tau$~Ceti system are on close-to-circular orbits 
and so they do not clear large annuli.
Using just the best-fit parameters of the five planets from \citet{Tuomietal2013}
results in stable disk particle orbits all the way down to 0.1~AU separation from the 
orbit of the outermost planet (at 1.35~AU). 
Figure~\ref{fig:diskorbits5pl} shows that 
stable disk orbits also exist in the gap between the outermost two planets.

\subsubsection{Simulations with an Additional Planet}

One way to constrain the inner disk edge at greater distances from the
star is to assume that there is an additional planet in the
system further from the host star and to estimate the properties of this hypothetical companion
based on the available data. Although this scenario was
not specifically tested in the work of \citet{Tuomietal2013}, the radial
velocity data sets could not be expected to be very sensitive to planets
with masses of roughly that of Neptune on longer period ($>$5~yr) orbits.

\begin{figure}
\centering
\includegraphics[scale=0.3,angle=-90]{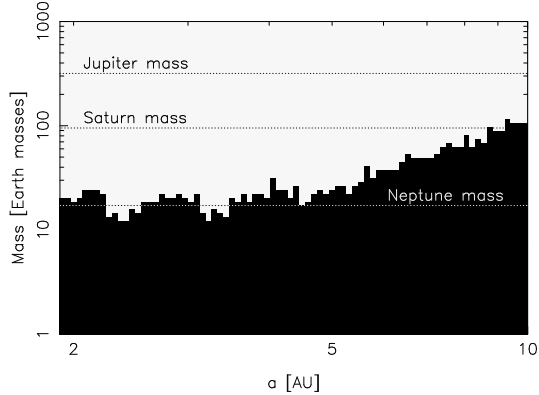}
\caption{
Black region shows planet mass (with the $\sim$30$^{\circ}$ system inclination taken into account)
and semimajor axis combinations that 
would not be
detectable using the analysis methods of \citet{Tuomietal2013}.
White region shows mass-semimajor axis combinations which cannot exist in the $\tau$~Ceti system
based on current RV data.
}
\label{fig:tuomi}
\end{figure}

In Figure~\ref{fig:tuomi}, we show the estimated detection threshold of additional
planets orbiting the star based on the radial velocity data analysis of \citet{Tuomietal2013}.
The area in the mass-period space where additional planet
candidates are ruled out (white area) has been
estimated by assuming there is an additional planet with a
semi-major axis in excess of those of the previously proposed
candidates. This estimation was performed by drawing a sample from
the posterior probability density of the parameters of the sixth planet
in a model that extends in the semi-major axis space from a minimum of
1.8~AU to a maximum value that we have chosen to be 10 AU,
the outermost bound of the inner disk edge consistent with the {\it Herschel} data. The
computations are performed as in \citet{Tuomietal2014}, and we have assumed
that the planetary eccentricity has a prior probability density that
penalizes high eccentricities as they approach unity,
because the eccentricities of low-mass planets appear to follow such a
distribution \citep{TuomiAngladaEscude2013}. 

It is worth noting that Jupiter-mass planets within 10~AU would have already
been detected by previous RV studies (e.g. \citealt{Pepeetal2011}, 
see also analysis by \citealt{Cummingetal2008}).
Using the analysis described above, Figure~\ref{fig:tuomi} shows that
we can push to lower planetary masses,
ruling out the existence of planets of Neptune-mass or larger within orbital
distances of 5~AU, and excluding the possibility that Saturn-mass or larger planets
exist in the system inside 10~AU.

We ran simulations with a sixth planet having twice Neptune's mass 
on circular orbits at 5-10~AU, 
and find that these systems are stable on long timescales (100~Myr).

While the disk could be cleared to large semimajor axes by a more eccentric outer planet
orbit, this situation would destabilize the inner five planets within millions of years at most.
Using a circular orbit for the outermost planet constrains the stable disk particle
orbits to just outside a few times the planet's Hill radius, as expected.  
While it is exciting to consider the possibility of additional planets in the system,
an additional planet with less than a Saturn mass (outside 5 AU) or less than a Neptune mass 
(inside 5 AU) would constrain the inner edge of the disk to larger distances from the star.
With the large uncertainties on inner edge of the disk, however, 
this additional layer of complexity is unwarranted. 
Future high-resolution images of the inner edge of $\tau$~Ceti's disk will provide much-needed
constraints on the actual location of the inner disk edge, and continued radial velocity observations
may detect additional, more distant planets.

\section{Discussion} \label{sec:disc}

\subsection{Properties of the $\tau$ Ceti Disk}

{\it Herschel} observations have confirmed the existence of the resolved debris disk originally
imaged by \citet{Greavesetal2004}, though the inclination we measure is 
$\sim$30$^{\circ}$ from face-on, which is different than the nearly edge-on alignment 
first reported from analysis of the SCUBA images.
Modelling the disk gives some weak constraints on the extent of the disk, which 
extends roughly 2-55~AU from $\tau$~Ceti.
There is no significant structure observed in the disk, and using a realistic dust grain 
spectrum (as opposed to blackbody) provided only moderately better constraints on the 
inner edge of the disk.  
Additional photometric or spectral data at far-IR wavelengths would be valuable
for constraining dust grain properties in this system.

\subsection{Disk Inner Edge}

The current data do not constrain the inner disk edge very strongly.
While an inner edge at 2-3 AU is consistent with both the SED- and image-based models, neither
can formally rule out a disk extending as close to the star as 1~AU, well into
the realm which may be populated by planets, or as far from the star as 10~AU,
allowing dynamical room for one or more additional planets.

We note that interferometric near-IR measurements have been made of the $\tau$~Ceti system
using the CHARA array \citep{diFolcoetal2007}.
They found they could reproduce the near-IR excess using 
a population of small ($<$1~$\mu$m) dust grains extending from the limits of their
observation field to extremely close to the star (3~AU to $\sim$0.1~AU), 
which overlaps with the region where the planets may exist.
However, the mass in dust grains is extremely small ($\sim$10$^{-9}$~$\Mearth$),
comparable to the mass of zodiacal (asteroidal) dust in our Solar System \citep{Hahnetal2002}.
Models show that dust produced by collisions at larger distances can inspiral 
(due to Poynting-Robertson drag) past planets in the inner solar system with little
disruption other than longer time spent inside planetary mean-motion resonances \citep{Nesvornyetal2011}.
For this reason, the presence of this tenuous dust in the inner portions of the $\tau$~Ceti
system does not rule out the planets.

The exquisite resolving power of ALMA should be able to image the inner edge of the main dust belt easily, 
and that will clarify which of three possibilities is true:
\begin{enumerate}
\item[1)] The disk extends well into the planetary regime ($<$1.35~AU), which would be
serious evidence against the planets proposed by \citet{Tuomietal2013}.
\item[2)] The disk edge is close to the orbit of the outermost planet 
($>$1.35~AU and $<$2.0~AU)
and the proposed five planet system is enough to constrain the disk edge.
\item[3)] The disk edge ends significantly far away from the outermost planet ($>$2.0~AU),
in which case another process must be invoked to explain the edge (e.g., another planet,
or possibly collisional processes).
\end{enumerate}
ALMA will also be more sensitive to larger dust grains that more closely trace the positions
of the parent bodies that are collisionally grinding to make 
the smaller dust grains observed at mid- and far-IR wavelengths.
The high-resolution of ALMA will also illuminate whether
the dust in the $\tau$~Ceti debris annulus is produced by a narrow ``birth ring'' as
has been observed in other debris disk systems 
\citep[e.g. AU~Mic;][]{Wilneretal2012,MacGregoretal2013}.

\subsection{The Disk-Planet Relationship} \label{sec:diskplanet}

If confirmed, $\tau$~Ceti's low-mass multiplanet system would fit with the results of 
simulations by \citet{Raymondetal2011} and
extend the trend observed by \citet{Wyattetal2012}: 
the presence of exclusively low-mass planetary systems ($<M_{\rm Saturn}$) 
and far-IR excess ($\sim$70~$\mu$m)
is strongly correlated for mature host stars.

These models hint that systems with planets of mass $>$~$M_{\rm Jup}$ are inherently unstable
in their early days \citep[e.g.][]{Raymondetal2012}.
We know from the structure of the Kuiper Belt that the four giant planets in our own 
Solar System have migrated,
and that a much more massive primordial Kuiper Belt is required to fuel this migration
\citep[e.g.][]{Gomesetal2005}.
It may be that the debris disk around solar analogue 
$\tau$~Ceti is brighter and more massive than the Kuiper Belt
because there are no giant planets in the system to migrate and disrupt the primordial planetesimal disk.

Unfortunately, the {\it Herschel} images do not provide very tight constraints on the
presence of gaps or clumps in the disk that may be due to perturbations by massive planets.

Resonant and secular perturbations by a planet on a disk can produce telltale clumps and gaps
in the dust disk, but the grain sizes that are most visible 
at 70~$\mu$m and 160~$\mu$m will be smeared out by radiation forces relative to 
the larger ($\sim$millimeter-sized) dust grains and parent planetesimals,
making these clumps much harder or even impossible to detect \citep{Wyatt2006}.
In addition, predicting clumps that may be present in the $\tau$ Ceti disk via 
numerical modelling of secular perturbations 
and mean-motion resonances in the planetesimal disk by the planets 
is not currently feasible.
These perturbations are
quite sensitive to the masses and exact orbits of the planets,
which have very large uncertainties or are even completely unconstrained,
as in the case of the angular orbital elements.

As shown in Section~\ref{sec:imagemodel}, the {\it Herschel} images
are best reproduced using a smooth disk model, with no structure.
However, because of the large beam size, this is not a very strong constraint 
on the smoothness of the disk.
A several AU-wide gap could easily be missed due to the {\it Herschel} resolution.
In order for a clump in the disk to be detectable, it would have to contain
more than $\sim$10\% of the total disk flux, 
based the sensitivities quoted in Table~\ref{tab:fluxes}.

In order to rule out clumps or gaps in the disk with any degree of certainty,
high-resolution, long wavelength observations are required in order to probe the 
distribution of large dust grains that are relatively unaffected by radiation forces.
Within the next few months, ALMA observations will be used to probe the inner portions of the
$\tau$~Ceti disk and provide some constraints both on the location of the inner edge of the disk
and on the smoothness of the disk, which in turn will provide limits on the mass and 
orbital properties of undetected massive planets in the system, independent of the RV data.

\section{Summary and Conclusions} \label{sec:concl}

$\tau$ Ceti hosts a bright debris disk that has been resolved by {\it Herschel}.  
The disk is uniform and symmetric, with a most likely inner edge at 2-3~AU 
(though inner edges 1-10~AU are not ruled out by the {\it Herschel} data) and an outer
edge at 55~$\pm$~10~AU.  
It is inclined from face-on by $35^{\circ}~\pm~10^{\circ}$ and can be fit by a surface density 
distribution of dust that increases linearly with distance from the star.

The proposed five planet system is not ruled out by the disk model, and our dynamical
simulations show that this system is stable for moderate planetary eccentricities.
If the outermost planet is what constrains the inner edge of the disk, the inner edge
should be at $\sim$1.5~AU.
If there is an additional, as-yet undetected planet (which is possible if its mass is below that of Neptune),
it could be constraining the inner disk radius farther away from the star.

It appears that there are no Jupiter-mass planets inside 10~AU in the $\tau$~Ceti system,
so the comparison to our Solar System may not be so appropriate.
If the proposed planets are real, the $\tau$~Ceti system is composed of
small rocky planets close to the star with a disk extending from the inner solar system 
out to Kuiper Belt-like distances from the star,
perhaps resembling our Solar System if the giant planets had failed to form and the
primordial planetesmial disk had not been disrupted by planet migration.
Future high resolution observations are required to constrain the edges of the disk, 
and to confirm the planetary system.

\section*{Acknowledgements}
The authors thank an anonymous referee for providing helpful comments on this paper.
S.M.L.\ and B.C.M.\ acknowledge an NSERC Discovery Accelerator Supplement which funded this work.
This work was also supported by the European Union through ERC grant number 279973 (G.M.K.)
M.B.\ acknowledges support from a FONDECYT Postdoctral Fellowship, project no.\ 3140479.


\end{document}